\documentclass[MSNbibl,nameyear,dvips]{arxstspdf}
\usepackage{flushend}
\usepackage{stfloats}

\volume{26}
\issue{2}
\pubyear{2011}
\firstpage{296}
\lastpage{298}
\doi{10.1214/11-STS349REJ}
\referstodoi{10.1214/10-STS349}

\begin{document}
\begin{frontmatter}

\title{Rejoinder}
\runtitle{Rejoinder}
\pdftitle{Rejoinder}

\begin{aug}
\author[a]{\fnms{Carl} \snm{Morris}\corref{}\ead[label=e1]{morris@stat.harvard.edu}}
\runauthor{C. Morris}

\affiliation{Harvard University}

\address[a]{Carl Morris is Professor of Statistics, Department of
Statistics, Harvard University, One Oxford Street,
Cambridge, Massachusetts 02138, USA \printead{e1}.
Ruoxi Tang obtained his Ph.D. from Harvard's
Statistics Department and is with a New York investment firm,
Bloomberg L.P., 731 Lexington Ave., New York, New York 10022, USA.}

\end{aug}



\end{frontmatter}

I thank all four discussants for their valuable insights.
   Before responding to their specific comments,
  let me help clarify to readers that adjustment for density (or
  likelihood, if appropriate) maximization
 is a method for approximation and not a stand-alone procedure for inference.
   The favorable
frequency pro\-perties of the ADM-SHP procedure rely particularly on the
flat prior chosen for the random effects variance $A$.
   After that, the responses of
 Partha Lahiri and Santanu Pramanik and of
Claudio Fuentes and George Casella are addressed.

   Because shrinkage factors $B_j$ are constrained to $[0,1]$,
 a Beta distribution ostensibly serves as a better approximation to the
 likelihood function or posterior density of  $B_j$  than does a Normal distribution.
    MLE and ADM methods are fitted based on computing two
derivatives, and they agree exactly when a Normal density
is chosen to approximate a~likelihood function (or a posterior with a flat prior).
  However, as Lahiri and Pramanik's Figure 2 shows,
sometimes no Normal distribution can closely approximate the distribution of a shrinkage
factor $B_j$ and then the MLE will yield misleading inferences
unless it can be liberated from its usual Normal approximation.

  ADM (Morris, \citeyear{Mor88}) was designed to approximate a given (one-dimensional) distribution with
 any chosen Pearson family, perhaps with a Normal distribution
 if for MLE purposes, or a Beta for shrinkage factors,
 or a Gamma, an Inverted Gamma, an $F$, a $t$, or a \mbox{\textit{Skew}-$t$} distribution
 for other situations.
   ADM does not alter a posterior density or a
likelihood function.
   The new curve that the ADM creates via multiplication by
the ``adjustment'' ($A$, in this paper) has no meaning
other than to provide a mode in the interior of the parameter
space that one believes will
lie closer to the mean of the actual density, or likelihood.

  The statistical properties of the ADM approximation depend
crucially on the corresponding properties of the procedure it
approximates.\vadjust{\goodbreak}
   While ADM can be used to approximate various
 Bayes procedures, for proper and for improper priors, that is not the goal in this paper.
  Rather, the objective is to
provide estimates of shrinkage factors via calculations similar to those of
MLE procedures that improve on the MLE for resulting inferences about  random effects.
   The flat prior on $A$ was chosen neither for Bayesian reasons nor
for subjective reasons, but because it leads to
Stein's harmonic prior (SHP) on the Level-I parameter vector $\theta$
and yields formal Bayes point estimators of the random effects with verified
and dominant mean squared error risks in the frequency sense.
   The paper provides additional strong  evidence that the formal Bayes posterior
intervals, whether computed exactly or as approximated by ADM,
meet (or nearly meet)  their nominal (95\% in the paper) confidence coverage rates
in the equal variance two-level Normal model, whatever be
the unknown between groups parameters $\beta, A$.

   Crucially, the conditional Level-II mean and variance of
   each random effect $\theta_j$
depends linearly on  $B_j$ and nonlinearly on $A$.
   For that reason ADM, which is designed to
   approximate a mean, starts in this application by approximating $B_j$
  with a Beta distribution, rather than
applying ADM directly to $A$ (perhaps with an approximating $F$ or
a Gamma distribution).
   By good fortune this turns out to be equivalent
   to setting $\hat{A}  = \operatorname{argmax}(A L(A)$)
with $L(A)$ the likelihood function [or perhaps a REML version of $L(A)$ if $r \geq
1$] so that $A$ legitimately can be viewed as a likelihood ``adjustment.''
   However, this adjustment actually arises as a principled choice based on three
considerations:
(a)~the established frequency properties of  formal Bayes procedures that  stem from SHP;
(b) the ADM approximation that uses a Beta distribution, for which the
   adjustment is $B_j (1-B_j)$;  and
(c) that the shrinkage factor $B_j$ enters linearly in the
first two Level-II moments, given ($\beta, A$),  of~$\theta_j$.

   Perhaps other confidence interval shrinkage procedures
for the Normal two-level model
have been proven to do  as well by  frequency standards
as  the procedures based on SHP and its ADM-SHP hybrid here.
We know from Figures 6 and 7 that coverage rates for these two procedures
hold up very well, even for very few groups.
  Data analysts regularly use two-level procedures
and report the nominal confidence interval coverage rates, but
it is unclear how often, if ever, the claimed frequency coverages
have been verified.

  We turn first to the comments of Lahiri and Pramanik.
  Analysts working with small area data, almost by definition of
  ``small,''
  encounter noisy estimates for individual small areas.
 Fortunately, SAE data sets
provide an opportunity to borrow strength by using information
from neighboring areas,
a technique for which the Fay--Herriot random effects mo\-del is widely used.
   However, maximizing the likelihood functions for the shrinkage factors
 that arise in such models not uncommonly produces full or nearly full shrinkages,
 as Lahiri and Pramanik show in Figures 1 and~2.
    In such cases MLE procedures typically
 (and non-conservatively) overestimate shrinkages and produce
  intervals  too narrow to meet their nominal confidences.
   That concern has inspired La\-hiri, with Pramanik and other co-authors,
to develop procedures that reduce or avoid over-shrinkage,
and ADM reasoning has helped them with that.

  The ADM  (dashed) curves in Figures 1 and 2  of Lahiri--Pramanik are
Beta densities that  show each state's own density
plotted  against $B_j = V_j/(A + V_j)$ (solid curves).
  If these densities were defined with respect to $dB_j$,
ADM then would have to be multiplied by $B_j(1-B_j)$,
the adjustment for a Beta density,
to produce a new curve with a mode aimed to lie
closer to the mean $E(B_j |$  data)  of the (``exact'')
posterior density (solid curve) of $B_j$ than does its own mode.
   However, that adjustment already has been made in Figures 1 and~2,
   and we see that all four dashed curves in Figure 2
   are maximized when $B_j < 1$
   [the Beta densities in Figures 1 and 2 are relative to the measure $dB_j/ (B_j(1-B_j))$].

    Lahiri and Pramanik ponder at the end of their first section whether ``\ldots\ there is
any need to find different adjustment factors, possibly depending on the $V_i,\ldots.$''
  Letting a prior depend on the available sample size means that it will
  change should more data become available.
  If  new data provide the only additional
information and their additional impact
is properly assessed in an updated analysis,
 there would be no basis for changing the prior.
   Perhaps this consideration will be useful even from a frequency perspective.

   Lahiri and Pramanik  ask,  ``How may the ADM
 method be useful in a non-Bayesian paradigm?,''
 describing  the SHP and ADM-SHP procedures as  ``essentially
 Bayesian.''
   That second section mainly concerns whether and how well ADM-like ideas
can help enhance familiar frequentist procedures such as EBLUP, REML,
MLE, and their own AML modification of ADM for estimating $A$ in the presence
of  unknown (nuisance) regression coefficients $\beta$.
   Their likelihood adjustment $g(A)$ is designed for the same two-level regression
  model as is the procedure in  Section 2.8 of the paper.

      They investigate likelihood adjustment factors\break
  other than $A$ by considering the resulting bias of shrinkage factors.
     A likelihood multiplier $A^{q}$  with  $0 < q < 1$
will increase shrinkages.
  These may be effective if $q$ is not too close to $0$
 ($q = 0$ returns us to MLE's problem of maximizing at the boundary).
   Such powers arise in our paper when ADM approximations
are developed for scale-invariant priors on $A$.
  There is little reason to consider $q > 1$ since the SHP rule ($q=1$) already
is quite conservative.
  With  $q < 1$ the resulting confidence interval estimators
may have insufficient coverage for some hyperparameters, particularly
for larger values of $A$.
   The bias of $\hat{B_j}$ may not provide the best criterion, as the
James--Stein shrinkage estimator is the uniformly minimum variance
 unbiased estimator of $B$  in
the equal variances setting, and then that unbiasedness comes at the cost of allowing the
shrinkage factor to exceed $1.00$,  making the James--Stein rule inadmissible.

  Lahiri and Pramanik's referring to the SHP and the ADM-SHP
procedures as  ``essentially Bayesian'' could suggest to  some
 frequentists that these rules are to be avoided.
  As already noted, Stein's (improper) harmonic prior has been chosen
here for the excellent frequency performance it endows on its
formal Bayesian point and interval  procedures.
   From a frequency perspective,
   any procedure that uniformly (whatever the unknown parameters)
    outperforms traditionally accepted frequentist procedures
     must be accepted, even preferred,
   regardless of how it has been or could be constructed.
   As is well known, and as Fuentes and Casella also emphasize, the fundamental
theorem of frequentist decision\vadjust{\eject} theory asserts that all admissible procedures
are essentially Bayesian, that is, are constructed from proper or formal priors.
   Procedures not thusly constructed can be improved upon uniformly.
   The ADM-SHP procedure here also performs well in repeated sampling,
and it too compares favorably with many procedures
regularly used by frequentists, with excellent confidence interval coverages.

    Claudio Fuentes and George Casella confine their
discussion to the equal variances case, even though real data
almost always  involve unequal variances.
    They have adopted this setting, as have many theorists, because
 the equal variance setting enables mathematical calculations
 which otherwise would be nearly intractable.

  Their discussion starts by considering shrinkage estimates of the vector $\theta$
that would arise if the Level-II variance $A$ were allowed to be
negative, showing that this inevitably leads to impossible distributions on $\theta$.
   We are reminded that the James--Stein estimator
   otherwise would be  admissible for $k \geq 3$, which would
   violate fundamental theorems in decision theory.
   The case $k=2$ is not considered, although then the James--Stein estimator reduces
to the unbiased sample mean vector, which is known to be admissible.

    Even so, being aware of what happens if integrals over $A$ (not $\theta$)
 are extended to include $-V \leq A < 0$ gives insight into
 the James--Stein estimator's over-shrinkage problem.
      It inspires the obvious
 and successful idea of  truncating at $A=0$, in which case
  the resulting flat prior on $A$ makes
  the likelihood function of $A$ agree with the posterior density and in turn this
  induces Stein's harmonic prior (SHP) on $\theta$.
   Extending the integral to allow $A < 0$ even
enables an easy gamma-function approximation to the SHP
shrinkage factor when $A$ is large, which reveals the
similarities between the SHP and the James--Stein shrinkage factors when
shrinkages are small.

  I  appreciate  Fuentes and Casella's
reminding readers that the ADM-SHP estimator is minimax
in the equal variance Normal setting, and for noting that the proof
is an immediate consequence of Al Baran\-chik's 1970 result.
  Their discussion about the left panel of Figure 1 embraces the range of
 minimax procedures covered by Baranchik's result.
  Al Baran\-chik was a Hunter College professor for over 40 years,
  after having been Charles Stein's Ph.D. student
and a colleague to many of us at Stanford when he proved his theorem for his 1964 dissertation.
  Al passed away not long ago, but ``Baranchik's  minimax theorem'' is forever.

  The right-hand panel of their Figure 1 plots risks
as a function of  $\theta$, revealing the SHP risk
to be uniformly lower than that of its ADM-SHP approximation.
 This must happen in the equal variance setting because
 the ADM approximation of  SHP's shrinkage factor
   always underestimates slightly, as seen in Figure 2 of Section~2.7.
      That makes ADM-SHP estimators of $\theta$ be more conservative than SHP estimators,
    which forfeits some of SHP's risk improvement over the sample mean vector.

  Fuentes and Casella point out that frequency
minimax theorems in the spirit of Stein estimation also
have been developed for non-Normal models settings.
  True, and the earliest non-Normal minimax   results
I~remember were for Poisson estimation,
by Clevenson and Zidek and  by J. T. Gene Hwang.
  However,  frequency confidence interval evaluations for  two-level
Poisson models largely have been ignored.
   In  practice, Bayesian methods are used for various non-Normal settings to
provide inferences in multilevel models that include
 posterior interval estimates for random effects.
    Again, there have been very
 few global evaluations to determine whether these
 Bayesian intervals can serve as approximate confidence
intervals as Level-II hyperparameters va\-ry throughout their range.

  Christiansen and Morris (\citeyear{ChrMor97})  used ADM to approximate
shrinkage factors for a two-level Poisson random effects regression model.
  Conjugate Gamma Level-II distributions are specified there
  to ensure existence of conditional
shrinkage factors.
   Just as here, the ADM approximation to the SHP shrinkages there
used Beta distribution approximations of shrinkage distributions to obtain
component-wise point and interval   estimates for the Poisson random effects.
  (The SHP is transported there to the Poisson setting via a shrinkage factor analogy.)
  The  results there have been implemented computationally by
  the PRIMM (Poisson regression interactive multilevel modeling) software.
     Our frequency-based evaluations of the resulting interval estimates
(limited to using PRIMM for simulation methods) successfully have met
frequency coverage standards, even for quite small $k$ and for
unequal sample sizes, regardless of  the hyperparameters tested.
    The PRIMM procedure can serve SAE with Poisson multilevel data,
  such as that of Manton, Woodbury and Stallard (\citeyear{ManWooSta81}).

  I extend special appreciation to Dr. Lahiri for inviting this paper
  and for organizing its discussion, in addition to
  his participating in the discussion.

\vspace*{-3pt}

\end{document}